\documentclass[journal,draftcls,onecolumn,12pt,twoside]{IEEEtran}
\normalsize

\ifCLASSINFOpdf

\usepackage{amsmath}
\usepackage{amsfonts}

\usepackage{amsmath}
\usepackage{amsfonts}
\usepackage{epsfig}
\usepackage{amssymb}
\usepackage{cite}
\usepackage{color,soul}
\usepackage{subfigure}
\usepackage{amsmath,amsthm,epsfig,amssymb,verbatim,amsopn,cite,multirow}
\usepackage{multirow}
\usepackage{rotating}
\usepackage{graphicx}
\usepackage{tabularx}
\usepackage{array}
\usepackage{graphicx,dblfloatfix}
\usepackage{blindtext}
\usepackage{ragged2e}
\usepackage{xcolor,colortbl}
\definecolor{Gray}{gray}{0.85}
\usepackage{amsthm,amssymb,amsmath,bm}
\usepackage{fancyhdr}
\usepackage[subfigure]{tocloft}
\usepackage[font={small}]{caption}
\usepackage{algorithmic}
\usepackage[Algorithm]{algorithm}

\begin{document}

\title{Information-Theoretic Security or Covert Communication}
\author{\IEEEauthorblockN{Moslem Forouzesh, Paeiz Azmi, \textit{Senior Member, IEEE,} Nader Mokari, \textit{Senior Member, IEEE,}  Kai Kit Wong, \textit{Fellow, IEEE}  and Dennis Goeckel,  \textit{Fellow, IEEE} }  \textsuperscript{}\thanks{\noindent\textsuperscript{} Moslem Forouzesh is with the Department of Electrical and
		Computer Engineering, Tarbiat Modares University, Tehran, Iran
		(e-mail: m.Forouzesh@modares.ac.ir).
		
		Paeiz Azmi is with the Department of ECE, Tarbiat Modares University,
		Tehran, Iran (e-mail: pazmi@modares.ac.ir).
		
		Nader Mokari is with the Department of ECE, Tarbiat Modares University,
		Tehran, Iran (e-mail: nader.mokari@modares.ac.ir).
		
		Kat-Kit Wong is with the Department of Electronic and Electrical Engineering, University College London, WC1E 7JE, United
		Kingdom (e-mail: kai-kit.wong@ucl.ac.uk).
		
		D. Goeckel is with the Electrical and Computer Engineering Department, University of Massachusetts, Amherst, Massachusetts (e-mail:
		goeckel@ecs.umass.edu).
		
		This work was supported in part by the  National Science Foundation under Grant CNS-1564067.
}}
\maketitle
\begin{abstract}
	Information-theoretic secrecy, in particular the wiretap channel formulation, provides protection against interception of a message by adversary Eve and has been widely studied in the last two decades.  In contrast, covert communications under an analogous formulation provides protection against even the detection of the presence of the message by an adversary, and it has drawn significant interest recently.  These two security topics are generally applicable in different scenarios;  however, here we explore what can be learned by studying them under a common framework.  Under a similar but not identical mathematical formulation, we introduce power optimization problems for each of the secrecy and the covert communications scenario, and we exploit common aspects of the problems to employ similar tools in their respective optimizations. \textcolor{black}{Moreover, due to the practical limitations, we assume only channel distribution information (CDI) are available for the secrecy and the covert communications scenarios.}  We then provide extensive numerical results to consider the performance of each of the schemes to understand the performance of each approach for various system parameters.  These results can be used, for example, to understand the difference in achievable rate that would be entailed in adopting a more restrictive covert communications approach rather than standard information-theoretic secrecy.
\end{abstract}
\section{Introduction}\label{Introduction}
Security is a major challenge in communication networks, and wireless communications links can make security even more challenging due to their broadcast nature.   Cryptographic security is the current state-of-the-art in practice, but, in recent years, information-theoretic security obtained by exploiting the physical layer has garnered a lot of attention, often employing the wiretap model introduced by Wyner\cite{Wyner}. In his pioneering work, Wyner demonstrated that if an eavesdropper's channel is a degraded version of the legitimate user's channel, then the legitimate user can achieve a positive information rate - known as the secrecy rate - at which the {\em content} of the message is kept confidential from the eavesdropper. 

More recently, there has been great interest in understanding the fundamental performance limits of techniques which aim to hide the {\em existence} of a communication from an observer; this has been termed ``covert'' communications.   The modern study of the limits of covert communications was initiated in \cite{korzhik05} and then addressed independently in \cite{bash_isit, bash_jsac} for additive white Gaussian noise (AWGN) channels;  these studies motivated significant further study \cite{jaggi_isit}\cite{Bloch}, \cite{Wang} to rapidly characterize the performance of covert communications over discrete memoryless channels (DMCs) and AWGN channels.  In \cite{jammer}, the authors study the capability of a source to transmit covertly and reliably to a destination with the help of a  jammer in the presence of the adversary. 

Critical to the security performance of a system is the knowledge at various participants about the characteristics of the multipath fading environment:  both the distribution of the channel fading, which we will term the channel distribution information (CDI), and the current instantiation of the channel fading, which is termed the channel state information (CSI).  There is a significant amount of work in information-theoretic secrecy done under the assumption of uncertain channel state information (CSI), while recently covert communication with uncertain CSI has been investigated in \cite{Covert} and \cite{J.Wang}.   A primary reason for this assumption is that both Eve and Bob are normally assumed to be passive and, if Bob feeds back the estimate of the channel between Alice and himself, Eve can damage this feedback signal by artificial noise,  which leads to low performance \cite{Jose} \cite{Zhou}.   While recent work in information-theoretic secrecy and covert communications has studied system design and performance under uncertain CDI in \cite{Abedi} and  \cite{Imperfect_CDI}, respectively, the standard is to assume perfect CDI as in  \cite{L.Sun}- \cite{Olabiyi} and \cite{Hu}, \cite{Xu},  respectively.  Indeed, in this paper, we assume that the CDI of the multipath fading channels is available to Alice, but we assume that the CSI of the channels from Alice to the eavesdropper (Eve) and Alice to the intended receiver (Bob) are unknown at the transmitter (Alice).

Information-theoretic security and covert communications are generally studied separately, because they have different goals and thus a selection is generally made based on the requirements of the application.  However, here we explore these two techniques together for two reasons:  (i) {\em Analytical:}   There might be commonalities in the optimization problems presented that allow us to employ similar techniques; and (ii)  {\em Application:}  It might be of interest to understand the cost of hiding the existence of the message (covert communications) rather than just its content (information-theoretic secrecy) in a given scenario.  
The system architecture considered will have the parties from the basic wiretap channel, with transmitter Alice, receiver Bob, and adversary Eve, but also with a jammer to (possibly) aid in secret or covert communication.   We will refer to this architecture as the wiretap channel throughout, with an understanding that a jammer has been added and that the adversary Eve in the covert case has a different goal than deciphering the message\footnote{Because the goal is different in the covert communications scenario, the adversary is often termed warden Willie, but we will use Eve as the adversary here in both scenarios.}.

This paper makes the following significant contributions:
\begin{enumerate}
	\item{For the considered wiretap channel architecture, the power allocations for maximizing the secrecy rate and covert rate are addressed:
		
		\begin{itemize}
			\item {To solve the optimization problem in the case of information-theoretic security, we first find a tight lower bound of the ergodic secrecy rate; then, we adopt the Successive Convex Approximation (SCA) method to maximize it.}
			\item {In order to solve the optimization problem for the covert communications scenario, we first find a tight lower bound of the ergodic covert rate; then, we employ an auxiliary variable to maximize it.}
			\item{We show that the lower bounds of ergodic secrecy rate and covert rate are tight.}
	\end{itemize}}
	\item 
	We obtain a closed-form optimal power threshold for Eve in the covert communication application.
	\item  We present numerical results demonstrating the performance of the two techniques for various operating conditions.
\end{enumerate}

The rest of this paper is organized as follows.  We present the system model in Section \ref{Sytem Model}. The optimization problems are developed for information-theoretic security and covert communication in Sections \ref{Physical layer security} and \ref{Covert communication}, respectively.  In Section \ref{Solution}, the solutions to the optimization problems are investigated. We provide numerical results and discussions in Section \ref{Simulation Results}. Finally, Section \ref{Conclusion} presents our conclusions.

\section{System Model}\label{Sytem Model}
The precise system model under consideration for each of the information-theoretic and covert communications scenarios will be given in successive sections; here, we give aspects that are common to both systems.  The model consists of  Alice, Bob, a jammer, and one Eve. Alice aims to send a private message to Bob, and the jammer broadcasts a jamming signal to help Alice.  The distance from Alice to Bob, Alice to Eve, jammer to Bob, and that from jammer to Eve are denoted by $d_{ab}$, $d_{ae}$, $d_{jb}$, and $d_{je}$, respectively. Moreover, we denote the channel coefficients between Alice and Bob, Alice and Eve, jammer and Bob,  and between jammer and Eve as $h_{ab}$, $h_{ae}$, $h_{jb}$, and $h_{je}$, respectively. 
\textcolor{black}{
	Per the previous section, we assume that channel distribution information (CDI) but not channel state information (CSI) is available to Alice.}

The channel coefficients are assumed to be circularly symmetric complex Gaussian random variables with zero mean and unit variance. We consider a discrete-time channel with $Q$ time slots, each with a length of $n$ symbols. The data signal transmitted by Alice and the jamming signal transmitted by the jammer in each time slot can be written as ${{\boldsymbol{x}}_b} = \left[ {x_b^1,x_b^2,\dots,x_b^n} \right]$ and  ${{\boldsymbol{x}}_j} = \left[ {x_j^1,x_j^2,\dots,x_j^n} \right]$, respectively.  As motivated by the work in \cite{jammer}, we assume Gaussian codebooks and a Gaussian jamming signal; hence, each of ${{\boldsymbol{x}}_b}$ and ${{\boldsymbol{x}}_j}$ is a vector of independently and identically distributed (i.i.d.) Gaussian random variables with zero mean and unit variance. 

Information-theoretic security and covert communication methods have different requirements, assumptions, and applications, which we review here before proceeding to the formal problem formulations in the next section.  
\begin{enumerate}
	\item{{\em Information-Theoretic Security:}
		\begin{itemize} 
			\item Assumptions: 
			\begin{enumerate}
				\item{Since Alice is known to be active, Eve is assumed to know the instantaneous values of the channel (CSI), having estimated such from prior observations.}
				\item{Eve knows the codebook used by the data transmitter (Alice) and legitimate receiver (Bob).}
			\end{enumerate} 
			\item Application:  Prevent Eve from extracting information of the private message; in other words, in this method, avoiding detection of the communication by Eve is not necessary while  avoiding information access by Eve is very important and should be guaranteed. 
		\end{itemize}
	}
	\item{{\em Covert Communication:}  
		\begin{itemize}
			\item Assumptions:
			\begin{enumerate}
				\item{Since Alice is no known to be active, Eve does not have CSI but has knowledge of the distribution of the channel (CDI).}
				\item{Eve does not know the codebook used by Alice and Bob.}
			\end{enumerate}
			\item Application:  Prevent Eve from detecting the presence of the communication.
	\end{itemize}}
\end{enumerate}

\section{information-theoretic security}\label{Physical layer security}

The  received signal at receiver $m$ (Bob or Eve) is given by
\begin{align}
{{\bf{y}}_m} = \frac{{\sqrt {{\left( {1 - \rho } \right){P_t}}} {h_{jm}}{{\boldsymbol{x}}_j}}}{{d_{jm}^{\alpha /2}}} + \frac{{\sqrt {{\rho P_t}} {h_{am}}{{\boldsymbol{x}}_b}}}{{d_{am}^{\alpha /2}}} + \boldsymbol{\eta}_m,
\end{align}
where $\rho$ and $P_t$ are the power allocation factor and total transmit power, respectively. Hence, $\left( {1 - \rho } \right){P_t}$ and $\rho P_t$ are the allocated power for the jamming signal and the transmission of Alice's message to Bob, respectively, $\alpha$ denotes the path-loss exponent, and $\boldsymbol{\eta}_m \sim \mathcal{CN}\left( {\bf{0}},\sigma _m^2 {\bf I}_n\right)$
represents the received noise vector at user $m$. Moreover,  ${\bf{I}}_n$
is the $n\times n$ identity matrix, $\bf{0}$ is an $n\times 1$ zero vector, and $\sigma _m^2$ is the variance of the noise at a given receiver.
Following \cite{R1} and  \cite{R2}, we assume Bob has knowledge of the predefined jamming signal and is able to cancel it. As a consequence, the signal-to-interference plus noise ratio (SINR) at Eve and Bob can be expressed as 
\begin{align}
{\rm{SINR}}_e ={\frac{{{\rho P_t}{{\left| {{h_{ae}}} \right|}^2}d_{ae}^{-\alpha}}}{{ {\left( {1 - \rho } \right){P_t}}{{\left| {{h_{je}}} \right|}^2}d_{je}^{-\alpha}+\sigma _e^2}},} \,\,\,\,\,
{\rm{SINR}}_b={\frac{{{\rho P_t}{{\left| {{h_{ab}}} \right|}^2}}}{{d_{ab}^\alpha \sigma _b^2}},}
\end{align}
respectively.
\begin{figure}[t]
	\centering
		\includegraphics[width=3.6in,height=3.2in]{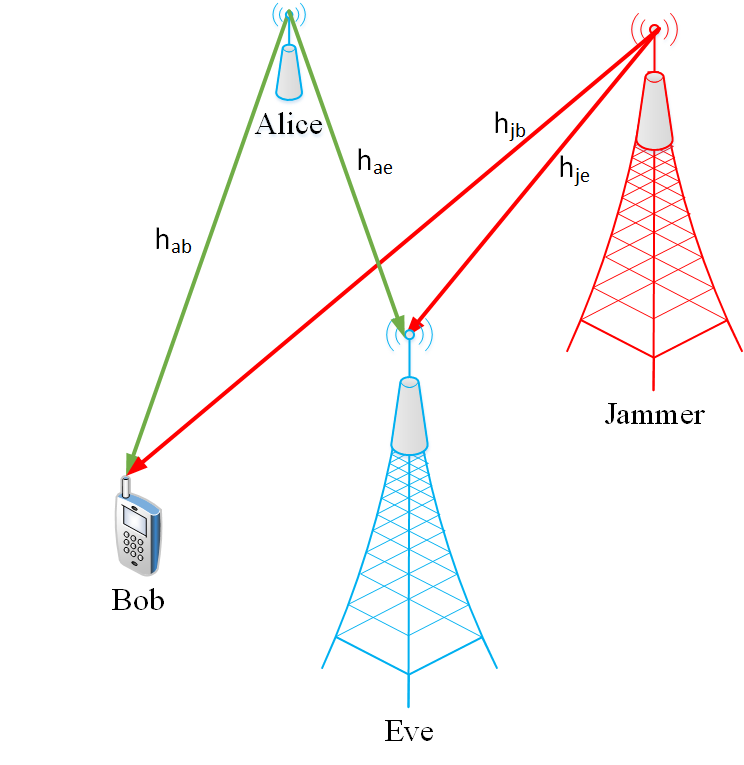}
	\caption{System model.}
	\label{Sys}
\end{figure}
Therefore, the instantaneous secrecy rate at  Bob can be written as \cite{SR}:
\begin{align}
R_{\sec } \left( \rho\right) = {\left[ {\log_2 \left( {1 +{\rm SINR}_b} \right) - \log_2 \left( {1 + {\rm SINR}_e} \right)} \right]^ + },
\label{secrecy_capacity}
\end{align}
where ${\left[ x \right]^ + }$ is defined as $\max \left\{ {x,0} \right\}$.

\subsection{Optimization Problem}\label{Optimization Problem_phy}

The system aims to maximize the average  secrecy rate at Bob subject to a transmit power constraint $P_t$. Hence, the optimization problem is formulated as follows:
\begin{subequations}\label{Opt_phy}
	\begin{align}
	&\max_{\rho} {\mathbb{E}\left\{\left[ {\log_2 \left( {1 +{\rm SINR}_b} \right) - \log_2 \left( {1 + {\rm SINR}_e} \right)} \right]^ + \right\}}\\& \,\,\mbox{s.t.}~~0\le\rho \le1 \label{powercon0},
	\end{align}
\end{subequations}
where $\mathbb{E} \left\{.\right\}$ is the expectation operator.

\section{Covert Communication}\label{Covert communication}
For the case of covert communication, Alice may be transmitting or she may not be, so there is a received signal model for each scenario.  In particular, the received signal at receiver $m$ (Bob or Eve) is 
\begin{equation}
{\boldsymbol{y}_m} = \left\{\begin{array}{ll}
{\frac{{\sqrt {{\left( {1 - \rho } \right){P_t}}} {h_{jm}}{\boldsymbol{x}_j}}}{{d_{jm}^{\alpha /2}}} + \boldsymbol{\eta}_m}, &\mbox{ }{{\Psi_0}}\\
{\frac{{\sqrt {{\rho P_t}} {h_{am}}{\boldsymbol{x}_b}}}{{d_{am}^{\alpha /2}}} +\frac{{\sqrt {{\left( {1 - \rho } \right){P_t}}} {h_{jm}}{\boldsymbol{x}_j}}}{{d_{jm}^{\alpha /2}}}+ \boldsymbol{\eta}_m}, &\mbox{ }{{\Psi_1}},
\end{array}\right.
\end{equation}
where $\Psi_0$ specifies the case where Alice does not transmit any message to Bob, while $\Psi_1$ states that Alice transmits a message to Bob.

Because Eve's CSI in unknown at Alice and the distribution of the multipath fading and noise are independent, the conditional distribution of each symbol of the received signal at Eve given $h_{ae}$ and $h_{je}$  follows $ y_m |h_{ae},h_{je} \sim {\cal C}{\cal N}\left( {0,\sigma _e^2 + \lambda } \right)$, where $\lambda$ is a  random variable which can be written as
\begin{align}
\lambda  = \left\{ {\begin{array}{*{20}{l}}
	{\frac{\left( {1 - \rho } \right){P_t}}{d_{je}^\alpha}{{\left| {{h_{je}}} \right|}^2},}&{ \,{\Psi _0}}\\
	{\frac{  \rho {P_t}}{d_{ae}^\alpha}{{\left| {{h_{ae}}} \right|}^2}+\frac{\left( {1 - \rho } \right){P_t}}{d_{je}^\alpha}{{\left| {{h_{je}}} \right|}^2},}&{\,{\Psi _1},}
	\end{array}} \right.
\end{align}
Since $h_{ae}$ and $h_{je}$ are circularly symmetric complex Gaussian with zero mean and unit variance and they are independent, the distribution of $\lambda$ is
\begin{align}\label{distribution}
{f_{{\Lambda _\Psi }}}\left( \lambda  \right) = {\rm{ }}\left\{ {\begin{array}{*{20}{l}}
	{\frac{1}{{\left( {1 - \rho } \right){P_t}d_{je}^{ - \alpha }}}{e^{ - \frac{{\lambda d_{je}^\alpha }}{{\left( {1 - \rho } \right){P_t}}}}},}&{{\mkern 1mu} {\Psi _0}}\\
	{\frac{1}{{{P_t}\left( {\rho d_{ae}^{ - \alpha } - \left( {1 - \rho } \right)d_{je}^{ - \alpha }} \right)}}\left[ {{e^{ - \frac{\lambda }{{\rho {P_t}d_{ae}^{ - \alpha }}}}} - {e^{ - \frac{\lambda }{{\left( {1 - \rho } \right){P_t}d_{je}^{ - \alpha }}}}}} \right],}&{{\mkern 1mu} {\Psi _1}.}
	\end{array}} \right.
\end{align}
The SINR at Bob is given by
\begin{equation}
{\rm SINR}_b=\left\{ \begin{array}{ll}
{0},& {{\Psi_0}}\\
{\frac{{{\rho P_t}{{\left| {{h_{ab}}} \right|}^2}}}{{d_{ab}^\alpha \sigma _b^2}}},& {{\Psi_1}}.
\end{array} \right.
\end{equation}
When Eve mistakenly decides $\Psi_1$ while $\Psi_0$ is true, a false alarm (FA) with probability $\mathbb{P}_{FA}$ occurs. Moreover, if Eve decides $\Psi_0$ while $\Psi_1$ is true, a missed detection (MD) with probability $\mathbb{P}_{MD}$ occurs. Alice and Bob achieve covert communication when, for $\varepsilon > 0$ \cite{bash_jsac}:
\begin{align}\label{decision}
\mathbb{P}_{MD}+\mathbb{P}_{FA}\ge 1-\varepsilon, \,\,\, \text{as}\,\,\, n \to \infty.
\end{align}
\textcolor{black}{We assume the  decision rule at Willie is the standard power detector,
	$
	\frac{{{Y_e}}}{n}	\mathop \gtrless\limits_{\Psi_0}^{\Psi_1} \vartheta,
	$
	that Willie is likely to employ in practice \cite{jammer}, \cite{Covert}, \cite{S. Lee}.}
Moreover, ${Y_e} = {\sum\limits_{\ell  = 1}^n {\left| {y_e^\ell } \right|} ^2}$ is defined as the total power received by Eve in a given time slot, and $\vartheta $ is the threshold for the decision at Eve.  Conditioned on $\lambda$, the FA and MD  probabilities are given by \cite{Covert}
\begin{align}\label{PFA}
&{\mathbb{P}_{FA}}(\lambda) = \mathbb{P}\left( {\frac{{{Y_e}}}{n} > \vartheta \left| \lambda, {{\Psi_0}} \right.} \right) = \mathbb{P}\left( {\left( {\sigma _e^2 + \lambda } \right)\frac{{\chi _{2n}^2}}{n} > \vartheta \left| \lambda, {{\Psi_0}} \right.} \right),\\
&{\mathbb{P}_{MD}}(\lambda) = \label{PMD}  \mathbb{P}\left( {\frac{{{Y_e}}}{n} < \vartheta \left| \lambda, {{\Psi_1}} \right.} \right) = \mathbb{P}\left( {\left( {\sigma _e^2 + \lambda } \right)\frac{{\chi _{2n}^2}}{n} < \vartheta \left| \lambda, {{\Psi_1}} \right.} \right),
\end{align}
where ${\chi _{2n}^2}$ is a  random variable with chi-squared distribution with $2n$ degrees of freedom.   Much of the recent work in covert communications \cite{bash_jsac} \cite{jaggi_isit}\cite{Bloch}, \cite{Wang} has focused on performance of the covert communications system as a function of the codeword length $n$.  However, in analogy to the insight provided by the outage approach to standard wireless communications, a number of authors have recently considered letting $n \rightarrow \infty$ and then considering the probability that channel conditions occur such that covert communications is achieved \cite{Covert}\cite{Imperfect_CDI}\cite{S.Ya}.  Here, we also take such an ``outage'' approach.

According to the strong law of large numbers (SLLN), $\frac{\chi _{2n}^2}{n}$  converges to 1, and, based on Lebesgue's dominated convergence theorem \cite{coverage}, we can rewrite \eqref{PFA} and \eqref{PMD} as:  
\begin{equation}\label{pfa}
\begin{aligned}
\mathbb{P}_{FA} (\lambda) = \mathbb{P}\left( {\sigma _e^2 + \lambda } > \vartheta \left| \lambda, {{\Psi_0}} \right . \right) 
= \left \{ \begin{array}{ll}
1, ~~~ & \lambda > \vartheta - \sigma_e^2 \\
0,& \mbox{else}  \end{array}
\right .
\end{aligned}
\end{equation}
\begin{equation}\label{pmd}
\begin{aligned}
\mathbb{P}_{MD} (\lambda) = \mathbb{P}\left( {\sigma _e^2 + \lambda } < \vartheta \left| \lambda, {{\Psi_1}} \right . \right) 
= \left \{ \begin{array}{ll}
1, ~~~ & \lambda < \vartheta - \sigma_e^2 \\
0,& \mbox{else}  \end{array}
\right .
\end{aligned}
\end{equation}

Employing the distribution of  $\lambda$ from \eqref{distribution} yields the unconditioned probability of false alarm and missed detection, respectively, as:
\begin{align}\label{int1}
{\mathbb{P}_{FA}} &= \int_{\vartheta  - \sigma _e^2}^\infty  {\frac{1}{{\left( {1 - \rho } \right){P_t}d_{je}^{ - \alpha }}}{e^{ - \frac{{\lambda d_{je}^\alpha }}{{\left( {1 - \rho } \right){P_t}}}}}d\lambda}  =\left\{ {\begin{array}{*{20}{l}}
	\begin{array}{l}
	{{e^{\frac{{ - \left( {\vartheta  - \sigma _e^2} \right)}}{{\left( {1 - \rho } \right){P_t}d_{je}^{ - \alpha }}}}},}\\	
	\end{array}&{ \,\vartheta  - \sigma _e^2 \ge 0}\\
	{1,}&{\,\vartheta  - \sigma _e^2 < 0,}
	\end{array}} \right.
\end{align}
\noindent and
\begin{align}\label{int2}
&{\mathbb{P}_{MD}}  =   \int_0^{\vartheta  - \sigma _e^2} {\frac{{\left[ {{e^{ - \frac{\lambda }{{\rho {P_t}d_{ae}^{ - \alpha }}}}} - {e^{ - \frac{\lambda }{{\left( {1 - \rho } \right){P_t}d_{je}^{ - \alpha }}}}}} \right]}}{{{P_t}\left( {\rho d_{ae}^{ - \alpha } - \left( {1 - \rho } \right)d_{je}^{ - \alpha }} \right)}}d\lambda}= \nonumber
\end{align}
\begin{align}
\left\{ {\begin{array}{*{20}{l}}
	{1 + \frac{{\left[ {\left( {1 - \rho } \right)d_{je}^{ - \alpha }{e^{ - \frac{{\left( {\vartheta  - \sigma _e^2} \right)}}{{\left( {1 - \rho } \right){P_t}d_{je}^{ - \alpha }}}}} - \rho d_{ae}^{ - \alpha }{e^{ - \frac{{\left( {\vartheta  - \sigma _e^2} \right)}}{{\rho {P_t}d_{ae}^{ - \alpha }}}}}} \right]}}{{\rho d_{ae}^{ - \alpha } - \left( {1 - \rho } \right)d_{je}^{ - \alpha }}},}&{\vartheta  - \sigma _e^2 \ge 0}\\
	{}&{}\\
	{}&{}\\
	{0,}&{\vartheta  - \sigma _e^2 < 0.}
	\end{array}} \right.
\end{align}

\noindent Combining  \eqref{int1} and \eqref{int2} yields:
\begin{align}\label{MD-FA0}
&{\mathbb{P}_{FA}} + {\mathbb{P}_{MD}}=\left\{ {\begin{array}{*{20}{l}}
	{1 + {e^{\frac{{ - \left( {\vartheta  - \sigma _e^2} \right)}}{{\left( {1 - \rho } \right){P_t}d_{je}^{ - \alpha }}}}} + \frac{U}{{\rho d_{ae}^{ - \alpha } - \left( {1 - \rho } \right)d_{je}^{ - \alpha }}},}&{\vartheta  - \sigma _e^2 \ge 0}\\
	{}&{}\\
	{1,}&{\vartheta  - \sigma _e^2 < 0.}
	\end{array}} \right.
\end{align}
where $U = {\left[ {\left( {1 - \rho } \right)d_{je}^{ - \alpha }{e^{ - \frac{{\left( {\vartheta  - \sigma _e^2} \right)}}{{\left( {1 - \rho } \right){P_t}d_{je}^{ - \alpha }}}}} - \rho d_{ae}^{ - \alpha }{e^{ - \frac{{\left( {\vartheta  - \sigma _e^2} \right)}}{{\rho {P_t}d_{ae}^{ - \alpha }}}}}} \right]}$.

\subsection{Optimal Threshold for Eve}\label{opt_tr}
Eve will select a decoding threshold that minimizes ${\mathbb{P}_{FA}} + {\mathbb{P}_{MD}}$.    If $\vartheta \leq \sigma_e^2$, Eve will always declare that Alice is transmitting regardless of the observation, and, hence, not surprisingly, we note from \eqref{MD-FA0} that ${\mathbb{P}_{FA}} + {\mathbb{P}_{MD}}=1$.  Hence, Eve will select a $\vartheta$ such that $\vartheta > \sigma_e^2$, meaning that we can find the optimal $\vartheta$ using the first line in \eqref{MD-FA0}.  It is shown in Appendix B that the optimal $\vartheta$ is given by
\begin{align}\label{therishold0}
{\vartheta ^*} = \left( {\frac{{\rho {P_t}d_{ae}^{ - \alpha }\left( {1 - \rho } \right){P_t}d_{je}^{ - \alpha }}}{{\left( {1 - \rho } \right){P_t}d_{je}^{ - \alpha } - \rho {P_t}d_{ae}^{ - \alpha }}}} \right)\ln \left( {\frac{{\left( {1 - \rho } \right){P_t}d_{je}^{ - \alpha }}}{{\rho {P_t}d_{ae}^{ - \alpha }}}} \right) + \sigma _e^2.
\end{align}
\subsection{Optimization Problem}\label{Optimization Problem}
Our goal is to maximize the covert rate
from Alice to Bob subject to the transmit power constraint and the covert
communication condition, i.e., \eqref{decision} is satisfied. Hence,
we consider the following optimization problem:
\begin{subequations}\label{Opt}
	\begin{align}
	\max_{\rho} &~\mathbb{P}_{\psi_1} \mathbb{E}\left\{\log_2 \left(1+{\frac{{{\rho P_t}{{\left| {{h_{ab}}} \right|}^2}}}{{d_{ab}^\alpha \sigma _b^2}}}\right)\right\},\label{Opti_probb}\\
	\mbox{s.t.} &~ \eqref{powercon0}\nonumber \\
	&\min_{\vartheta} \left ( {\mathbb{P}_{FA}} + {\mathbb{P}_{MD}} \right ) > 1-\varepsilon.\label{covertcon}
	\end{align}
\end{subequations}
where $\mathbb{P}_{\psi_1}$ it the probability that Alice transmits.  
For solving \eqref{Opt}, we first solve $\mathop {\min }\limits_\vartheta  \;{\mathbb{P}_{FA}} + {\mathbb{P}_{MD}}$ to obtain the optimal $\vartheta$, denoted by $\vartheta ^*$, for Eve. Then we solve \eqref{Opt} based on $\vartheta^*$. 


\section{Solutions to the Optimization Problems}\label{Solution}
Problems \eqref{Opt_phy} and \eqref{Opt} are nonconvex: in \eqref{Opt_phy}, the objective function is not concave, and, in \eqref{Opt}, constraint \eqref{covertcon} is not convex. In the following, we propose novel algorithms to solve these optimization problems.

\subsection{The Information-Theoretic Security Case}\label{TIFS}
Per \eqref{secrecy_capacity}, the instantaneous secrecy rate is 
\begin{align}\label{ISR}
{R_{\sec }}\left( \rho  \right) = {\left[ {{{\log }_2}\left( {\frac{{1 + {\rm{SIN}} {{\rm{R}}_b}}}{{1 + {\rm{SIN}}{{\rm{R}}_e}}}} \right)} \right]^ + }.
\end{align}

We know that ${{\log }_2}\left( {\frac{{1 + {\rm{SIN}} {{\rm{R}}_b}}}{{1 + {\rm{SIN}}{{\rm{R}}_e}}}} \right)\le {\left[ {{{\log }_2}\left( {\frac{{1 + {\rm{SIN}} {{\rm{R}}_b}}}{{1 + {\rm{SIN}}{{\rm{R}}_e}}}} \right)} \right]^ + }$, hence,  $\mathbb{E}\left\{{{\log }_2}\left( {\frac{{1 + {\rm{SIN}} {{\rm{R}}_b}}}{{1 + {\rm{SIN}}{{\rm{R}}_e}}}} \right)\right\}\le \mathbb{E}\\ \left\{ \left[ {{{\log }_2}\left( {\frac{{1 + {\rm{SIN}} {{\rm{R}}_b}}}{{1 + {\rm{SIN}}{{\rm{R}}_e}}}} \right)} \right]^ + \right\}.$ In order to simplify the formulation  and solve the corresponding optimization problem, we maximize $\mathbb{E}\left\{{{\log }_2}\left( {\frac{{1 + {\rm{SIN}} {{\rm{R}}_b}}}{{1 + {\rm{SIN}}{{\rm{R}}_e}}}} \right)\right\}$ instead of $\mathbb{E}\left\{ {\left[ {{{\log }_2}\left( {\frac{{1 + {\rm{SIN}} {{\rm{R}}_b}}}{{1 + {\rm{SIN}}{{\rm{R}}_e}}}} \right)} \right]^ + }\right\}$; then, we show that the maximization of this lower bound is very tight by numerical results.
By exploiting this lower bound, the optimization problem \eqref{Opt_phy} is reformulated as:
\begin{align}
&\mathbb{E}\left\{R_{\sec } \left( \rho\right)\right\}\ge\mathbb{E} \left\{ { - \log_2 \left( {1 + \frac{{\rho {P_t}{{\left| {{h_{ae}}} \right|}^2}d_{ae}^{ - \alpha }}}{{\left( {1 - \rho } \right){P_t}{{\left| {{h_{je}}} \right|}^2}d_{je}^{ - \alpha } + \sigma _e^2}}} \right)} \right.  + \left. {\log_2 \left( {1 + \frac{{\rho {P_t}{{\left| {{h_{ab}}} \right|}^2}}}{{d_{ab}^\alpha \sigma _b^2}}} \right)} \right\}\nonumber
\end{align}

\begin{align}\label{ESR}
\begin{array}{l}
= \frac{1}{{\ln 2}}\left\{ {\underbrace {\mathbb{E}\left\{ {\ln \left( {1 + \frac{{\rho {P_t}{e^{\ln \left( {{{\left| {{h_{ab}}} \right|}^2}} \right)}}}}{{d_{ab}^\alpha \sigma _b^2}}} \right)} \right\}}_A - } \right. \underbrace {\mathbb{E}\left\{ {\ln \left( {\sigma _e^2 + \left( {1 - \rho } \right){P_t}d_{je}^{ - \alpha }{{\left| {{h_{je}}} \right|}^2} + \rho {P_t}d_{ae}^{ - \alpha }{{\left| {{h_{ae}}} \right|}^2}} \right)} \right\}}_B\\\,\,\,\,\,+
\left. {\underbrace {\mathbb{E}\left\{ {\ln \left( {\sigma _e^2 + \left( {1 - \rho } \right){P_t}d_{je}^{ - \alpha }{e^{\ln \left( {{{\left| {{h_{je}}} \right|}^2}} \right)}}} \right)} \right\}}_C} \right\}.
\end{array}
\end{align} 

In the following, we again will find a lower bound, this time of \eqref{ESR}, and then maximize this lower bound.  The numerical results will again illustrate that this lower bound is near to the ergodic secrecy rate.
We will lower bound $A$ in \eqref{ESR} as
\begin{align}\label{LA}
A=& \mathbb{E}\left\{ {\ln \left( {1 + \frac{{\rho {P_t}{e^{\ln \left( {{{\left| {{h_{ab}}} \right|}^2}} \right)}}}}{{d_{ab}^\alpha \sigma _b^2}}} \right)} \right\} \mathop  \ge \limits^{(a)} \ln \left( {1 + \frac{{\rho {P_t}{e^{\mathbb{E}\left\{ {\ln \left( {{{\left| {{h_{ab}}} \right|}^2}} \right)} \right\}}}}}{{d_{ab}^\alpha \sigma _b^2}}} \right).
\end{align}
Moreover, $B$ can be upper bounded by
\begin{align}\label{LB}
&B = \mathbb{E}\left\{ {\ln \left( {\sigma _e^2 + \left( {1 - \rho } \right){P_t}d_{je}^{ - \alpha }{{\left| {{h_{je}}} \right|}^2} + \rho {P_t}d_{ae}^{ - \alpha }{{\left| {{h_{ae}}} \right|}^2}} \right)} \right\}\nonumber\\&\mathop  \le \limits^{(b)} \ln \left( {\sigma _e^2 + \left( {1 - \rho } \right){P_t}d_{je}^{ - \alpha }\mathbb{E}\left\{ {{{\left| {{h_{je}}} \right|}^2}} \right\} + \rho {P_t}d_{ae}^{ - \alpha }\mathbb{E}\left\{ {{{\left| {{h_{ae}}} \right|}^2}} \right\}} \right),
\end{align}
and $C$ can be lower bounded by
\begin{align}\label{LC}
C =& \mathbb{E}\left\{ {\ln \left( {\sigma _e^2 + \left( {1 - \rho } \right){P_t}d_{je}^{ - \alpha }{e^{\ln \left( {{{\left| {{h_{je}}} \right|}^2}} \right)}}} \right)} \right\} \mathop  \ge \limits^{(c)} \ln \left( {\sigma _e^2 + \left( {1 - \rho } \right){P_t}d_{je}^{ - \alpha }{e^{\mathbb{E}\left\{ {\ln \left( {{{\left| {{h_{je}}} \right|}^2}} \right)} \right\}}}} \right),
\end{align}
In \eqref{LA}, \eqref{LB}, and \eqref{LC}, $(a), (b)$, and $(c)$ are obtained from Jensen’s inequality. Note that the functions $\ln \left( {1 + s{e^x}} \right)$ and $\ln \left({1+dx}\right)$ are convex for $s>0$ and concave for $d\in \mathbb{R}$, respectively, where $\mathbb{R}$ is the set of real numbers.

Next, our aim is to calculate ${\mathbb{E}\left\{ {\ln \left( {{{\left| {{h_{ab}}} \right|}^2}} \right)} \right\}}$ and ${\mathbb{E}\left\{ {\ln \left( {{{\left| {{h_{je}}} \right|}^2}} \right)} \right\}}$.  First:
\begin{align}
&\mathbb{E}\left\{ {\ln \left( {{{\left| {{h_{ab}}} \right|}^2}} \right)} \right\} = \int\limits_0^\infty  {\ln \left( x \right)} {f_{{{\left| {{h_{ab}}} \right|}^2}}}(x)dx = \int\limits_0^\infty  {\ln \left( x \right)} {e^{ - x}}dx \nonumber\\&= \left\{ {Ei\left( { - x} \right) - {e^{ - x}}\ln \left( x \right)} \right\}\mid _0^\infty= -\sum\limits_{k = 1}^\infty  \left( {\frac{1}{k} - \ln \left( {\frac{{k + 1}}{k}} \right)} \right) =- \gamma,
\end{align} 
where ${f_{{{\left| {{h_{ab}}} \right|}^2}}}(x)$ is the probability density function (pdf) of the random variable ${{\left| {{h_{ab}}} \right|}^2}$. Moreover, $\gamma$ is the Euler-Mascheroni constant with value $\gamma \simeq 0.577216$.
Similar to ${\mathbb{E}\left\{ {\ln \left( {{{\left| {{h_{ab}}} \right|}^2}} \right)} \right\}}$, the expectation of  ${ {\ln \left( {{{\left| {{h_{je}}} \right|}^2}}   \right)}}$ can be calculated, and, according to initial assumptions $ {\mathbb{E}\left\{ {{{\left| {{h_{je}}} \right|}^2}} \right\}}={\mathbb{E}\left\{ {{{\left| {{h_{ae}}} \right|}^2}} \right\}}=1$. Finally, the lower bound of the objective function can be rewritten as:
\begin{align}\label{obj_lb}
R_{lb}\left( \rho\right)=&\frac{1}{{\ln 2}}\left\{ {\ln \left( {1 + \frac{{\rho {P_t}{e^{ - \gamma }}}}{{d_{ab}^\alpha \sigma _b^2}}} \right) } \right. -
\ln \left( {\sigma _e^2 + \left( {1 - \rho } \right){P_t}d_{je}^{ - \alpha } + \rho {P_t}d_{ae}^{ - \alpha }} \right)\\& +
\left. {\ln \left( {\sigma _e^2 + \left( {1 - \rho } \right){P_t}d_{je}^{ - \alpha }{e^{ - \gamma }}} \right)} \right\}\nonumber
\end{align}
Hence, we reformulate  the optimization problem \eqref{Opt_phy} as follows:
\begin{align}\label{Opt_lb}
&\max_{\rho} {R_{lb}\left( \rho\right)}\\& \,\,\mbox{s.t.}~~0 \leq \rho \leq 1 \nonumber,
\end{align}
Note that the objective function is non-concave. As a result, the problem is non-convex. In order to solve this problem, we employ the Successive Convex Approximation (SCA) method, which converts the optimization problem to a convex one.  In particular,  \eqref{obj_lb} has the difference of two concave functions, and thus we adopt the difference of concave functions (DC) method to approximate it as a concave function. To this end, we rewrite \eqref{obj_lb}  as
\begin{equation}
\Xi \left( \rho\right) = \Im \left(\rho\right)-\Phi \left(\rho\right),
\end{equation}
where
\begin{equation}
\left\{\begin{aligned}
\Im \left( \rho  \right) = &
\frac{1}{{\ln 2}}\left\{ {\ln \left( {\rho {P_t}{e^{ - \gamma }} + d_{ab}^\alpha \sigma _b^2} \right)} \right. + \\&
\left. {\ln \left( {\sigma _e^2 + \left( {1 - \rho } \right){P_t}d_{je}^{ - \alpha }{e^{ - \gamma }}} \right) - {\ln}\left( {d_{ab}^\alpha \sigma _b^2} \right)} \right\}.\\\\
\Phi \left( \rho  \right) =& \frac{1}{\ln2}{\ln}\left( {\sigma _e^2 + \left( {1 - \rho } \right){P_t}d_{je}^{ - \alpha } + \rho {P_t}d_{ae}^{ - \alpha }} \right).
\end{aligned}\right.
\end{equation}
By employing a linear approximation, $\Phi \left(\rho\right)$ can be rewritten as 
\begin{align}
\Phi \left({\rho}\right) \simeq \tilde \Phi \left({\rho}\right) = \nonumber	\Phi \left({\rho}\left({s}-1\right)\right) 
+ {\nabla ^T}\Phi \left({\rho}\left({s}-1\right)\right)\left( {{\rho} - {\rho}({s}  - 1)} \right),
\end{align}	
where  $\nabla$ is the gradient operator. Therefore, we have
\begin{align}
\nabla \Phi \left( {\rho \left( {s - 1} \right)} \right) = \frac{1}{{\ln (2)}} \times \left[ {  \frac{{{-P_t}d_{je}^{ - \alpha } + {P_t}d_{ae}^{ - \alpha }}}{{\sigma _e^2 + \left( {1 - \rho } \right){P_t}d_{je}^{ - \alpha } + \rho {P_t}d_{ae}^{ - \alpha }}}} \right].
\end{align}
Finally, after utilization of the DC approximation, \eqref{Opt_lb} can be approximated to a convex optimization problem as follows
\begin{align}\label{final_ITS}
\max_{\rho, } &~\Im \left(\rho\right)-\tilde \Phi \left(\rho\right)\\
\mbox{s.t.} &~0 \leq \rho \leq 1,\nonumber
\end{align}
For the solution of this convex optimization problem \eqref{final_ITS} , we can
use available software such as the CVX solver \cite{CVX}.  The iterative algorithm employed is shown as Algorithm I. This algorithm is stopped when $\left| {\rho\left( {s  + 1} \right) - \rho \left( s  \right)} \right| \le \omega  $ is satisfied, where $\omega$ is the stopping threshold.

\begin{algorithm}[t]
	\caption{ITERATIVE POWER ALLOCATION ALGORITHM} \label{alg1}
	\begin{algorithmic}[1]
		\STATE  \nonumber
		Initialization: Set $s =0
		\left( {s \text{\hspace{.2cm}is the iteration number}} \right)$
		and initialize to $\rho(0)$.
		\STATE \label{set}
		Set $\rho=\rho(s)$ 
		\STATE  		
		Solve \eqref{final_ITS} and set  the result  to $\rho(s +1)$
		\STATE 
		If $\left| {\rho\left( {s  + 1} \right) - \rho \left( s  \right)} \right| \le \omega  $\\
		stop,\\
		else\\
		set $s = s + 1$ and go back to step \ref{set}
	\end{algorithmic}
\end{algorithm}

\subsection{The Covert Communication Case}

By substituting \eqref{therishold0}  into \eqref{int1} and \eqref{int2}, and since $\vartheta ^* \ge \sigma _e^2$ is always true, we can rewrite \eqref{Opt} as follows:

\begin{subequations}\label{Opt_c}
	\begin{align}
	\max_{\rho} &~\mathbb{P}_{\psi_1} \mathbb{E}\left\{\log_2 \left(1+{\frac{{{\rho P_t}{{\left| {{h_{ab}}} \right|}^2}}}{{d_{ab}^\alpha \sigma _b^2}}}\right)\right\},\label{Opt_probb}\\
	\mbox{s.t.} &~ \eqref{powercon0}\nonumber \\
	&\ln \left( {\frac{{\left( {1 - \rho } \right){P_t}d_{je}^{ - \alpha }}}{{\rho {P_t}d_{ae}^{ - \alpha }}}} \right)\frac{{\left( {1 - \rho } \right){P_t}d_{je}^{ - \alpha }}}{{\rho {P_t}d_{ae}^{ - \alpha } - \left( {1 - \rho } \right){P_t}d_{je}^{ - \alpha }}} \le \ln \left( \varepsilon  \right)
	\label{covertcon_c}.
	\end{align}
\end{subequations}
Similar to Section \ref{TIFS}, the lower bound of \eqref{Opt_probb} can be calculated.
By utilizing an auxiliary variable $t$, the optimization problem \eqref{Opt_c} is equivalent to the following problem
\begin{subequations}\label{Opt_e}
	\begin{align}
	\max_{\rho} &~\mathbb{P}_{\psi_1}\log_2 \left(1+{\frac{{{\rho P_t}{e^{-\gamma}}}}{{d_{ab}^\alpha \sigma _b^2}}}\right),\label{Opte_probb}\\
	\mbox{s.t.} &~ \eqref{powercon0}\nonumber \\
	& \left( {1 - \rho } \right){P_t}d_{je}^{ - \alpha }\ln \left( {\frac{{\left( {1 - \rho } \right){P_t}d_{je}^{ - \alpha }}}{{\rho {P_t}d_{ae}^{ - \alpha }}}} \right) - t\ln \left( \varepsilon  \right) \le 0\\
	&\rho {P_t}d_{ae}^{ - \alpha } - \left( {1 - \rho } \right){P_t}d_{je}^{ - \alpha } \le t
	\label{covertcon_e}
	\end{align}
\end{subequations}
The optimization problem \eqref{Opt_e} is convex; hence, for solving  \eqref{Opt_e}, we can use available software such as the CVX solver \cite{CVX}.


\section{Numerical Results}\label{Simulation Results}
\textcolor{black}{
	In this section, we present numerical results to evaluate the performance in the proposed information-theoretic security and covert communication scenarios.
	The considered simulation setting is listed in Table I.}
\begin{table}[b] 
	\centering
	\caption{ Simulation setting}
	\begin{tabular}{|m{2 em} |m{20 em} |m{4.5 em}|}  	
		
		\hline
		$d_{jb}$ & Distance between jammer and bob & \hspace{.55cm} $5$ m \\
		\hline
		$\alpha$ &   Path-loss exponent &\hspace{.55cm} $4$\\
		\hline
		\vspace{.2cm}
		$ \mathbb{P}_{\psi_1}$ & Probability of data transmission  by Alice & \hspace{.55cm}$0.5 $\\
		\hline
		$1-\varepsilon$  & Lower bound
		of detection error probability at Eve    & \hspace{.001cm}$0.9$, $0.99$, and $0.999$ \\
		\hline
		\textcolor{black}{$\sigma_b^2$}  & \textcolor{black}{Received noise power at Bob}  &\textcolor{black}{$-30$ dBW} \\
		\hline
		\textcolor{black}{$\sigma_e^2$ } & \textcolor{black}{Received noise power  at Eve}  &\textcolor{black}{$-30$ dBW }\\
		\hline
	\end{tabular}
	\label{table1}
\end{table}

\begin{table*}[h] 
	\begin{center}
		\caption{ Assumption and applications of covert communication and information-theoretic security}
		\begin{tabular}{|m{12 em} |m{14 em} |m{16.45 em}|} 
			\hline \rowcolor{red}	
			\hspace{1.5cm} Items & \hspace{.9cm}  Covert communication & \hspace{.8cm} Information-theoretic security  \\ [0.5ex] 
			\hline\hline
			Assumptions  &  The code-book is unknown at Eve  &   The code-book is known at Eve \\ 
			\hline
			Applications & To prevent detection of communication & To prevent extraction of private information   \\
			\hline
			Existence of legitimate jammer &   Covert rate increase &  Secrecy rate increase
			\\
			\hline
			Existence of Eve jammer & Covert rate increase  & Secrecy rate decrease\\
			\hline
		\end{tabular}
	\end{center}
	\label{table}
\end{table*}

\begin{figure}[t]
	\centering
	\includegraphics[width=4.2in,height=3.3in]{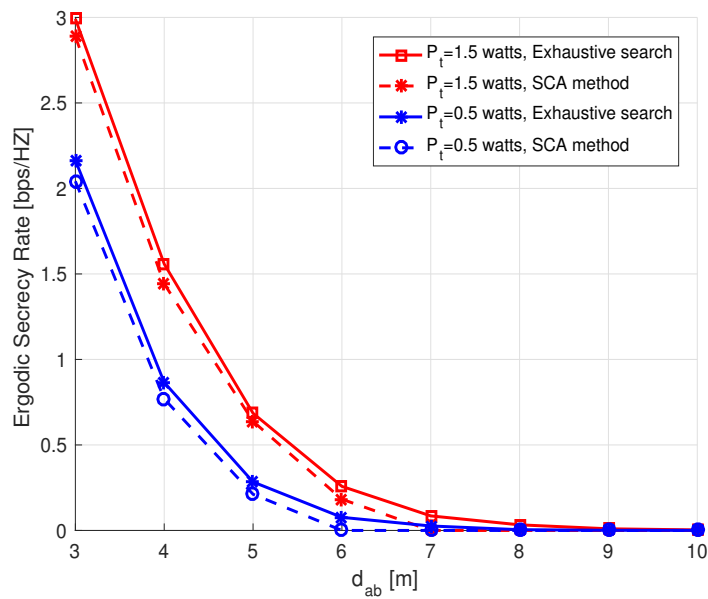} 
	\caption{\textcolor{black}{Ergodic secrecy rate \emph{vs} $d_{ab}$ and evaluation of the impact  of the total transmit power, $d_{ae}=5$ m(meter), $d_{je}=5$ m.}}
	\label{d_abs}
\end{figure}

\textcolor{black}{
	Furthermore, we summarize the considered assumptions and applications for covert communication and information-theoretic security methods in  Table II.
	As seen in this table, in information-theoretic security there is two cases 1) When the legitimate nodes transmit a jamming signal which leads to an increase in the secrecy rate, 2)  when Eve transmits a jamming signal which leads to a decrease in the secrecy rate  \cite{IST}. In covert communication, whether Eve or legitimate nodes transmit the jamming signal, the covert rate increases because the message is hidden in  the jamming signal \cite{jammer}.}

\textcolor{black}{
	Fig.~\ref {d_abs} shows the ergodic secrecy  rate versus the distance between Alice and Bob. Moreover, this figure compares the proposed solutions in the information-theoretic security scenario with optimal values which are obtained from the exhaustive search method.
	To study the tightness of the proposed lower bound on the secrecy  rate, we directly  solve the original optimization problem \eqref{Opt_phy} by the exhaustive search method.
	The proposed solution by SCA has about an 11.8\% performance gap with that from the exhaustive search.}

\begin{figure}[t]
	\centering
	\includegraphics[width=4.2in,height=3.3in]{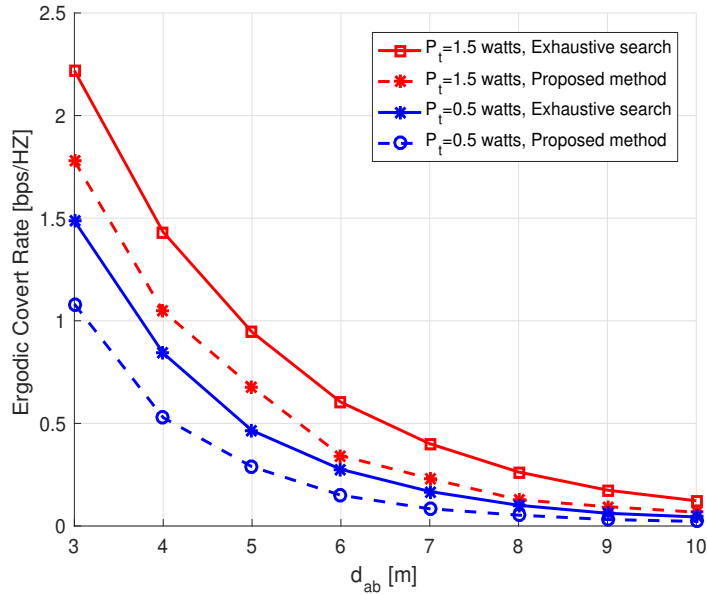} 
	\caption{\textcolor{black}{ Ergodic covert rate \emph{vs} $d_{ab}$ and evaluation of the impact of the total transmit power, $\varepsilon=0.1$, $d_{ae}=5$ m, $d_{je}=5$ m.}}
	\label{d_abc}
\end{figure}
Fig.~\ref{d_abc} shows the ergodic covert  rate versus the distance from Alice to Bob.  Moreover, this figure illustrates the impact of different values of the total transmit power on the ergodic covert rate. As seen, the ergodic covert rate is an increasing function with respect to the total transmit power. The reason is that the emitted jamming signal only affects the received signal at Eve and has no effect on Bob because he is able to cancel the jamming signal. Moreover, the proposed solution has approximately a 23.7\% performance gap with that obtained from the exhaustive search. A comparison between exhaustive search in Fig.~\ref{d_abs} and Fig.~\ref{d_abc} shows when Bob is near to Alice, the information-theoretic security scheme has higher rate than that of the covert communication scheme, but, when Bob is farther from Alice, the covert communication method has higher rate compared to the information-theoretic security method. This is at first surprising given the seemingly higher security requirement of covert communications. But recall that Alice and Bob share a secret codebook in the covert communications scenario, yielding a potential advantage. Mathematically, the reason it happens is that the covert communication requirement  \eqref{covertcon_c} and $d_{ab}$ are independent, and the sensitivity of \eqref{Opt_probb} to $d_{ab}$ is greater than the sensitivity of \eqref{Opt_phy} to $d_{ab}$ because of the difference in the two achievable data rates. In   information-theoretic secrecy,  when Bob gets further away  from Alice the  achievable rate  of the Alice-to-Bob channel approaches the achievable rate of the Alice-to-Eve channel which leads to zero secrecy rate, while in the covert communication case the covert rate at high values of $d_{ab}$ ($7-10$ m) is non- zero.

\begin{figure}[t]
	\centering
	\includegraphics[width=4.2in,height=3.3in]{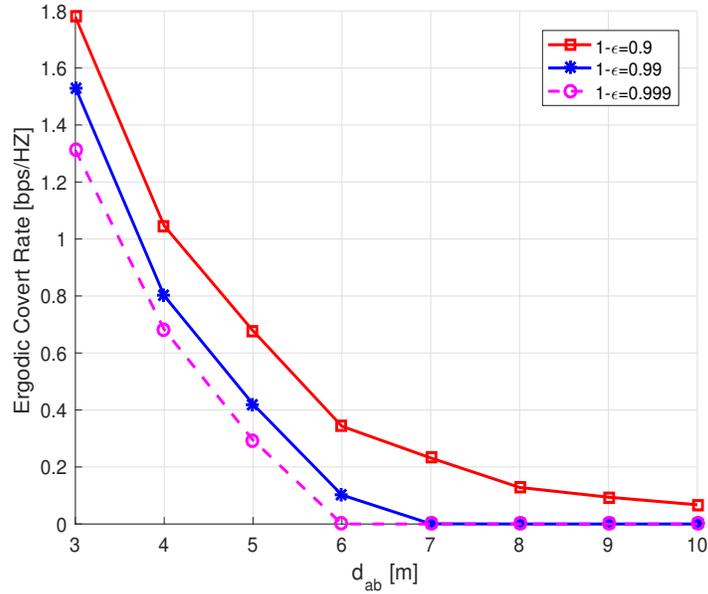} 
	\caption{ \textcolor{black}{Ergodic covert rate \emph{vs} $d_{ab}$ and evaluation of the impact of the lower bound of detection error probability at Eve on covert rate, i.e., $1-\varepsilon$, $P_t=1.5$ Watts, $d_{ae}=5$ m, $d_{je}=5$ m.}}
	\label{d_epsil}
\end{figure}
Fig.~\ref{d_epsil} depicts  the covert rate versus  $d_{ab}$ and evaluates the impact of the lower bound of detection error probability at Eve on the covert rate. As expected, increasing the lower bound of the detection error probability decreases the covert rate. 
As shown, if we aim to guarantee at least 99\% and 99.9\% detection error probability at Eve instead of 90\%, the covert rate decreases 1.16 and 1.76 times, respectively. By increasing the lower bound of the detection error probability, Alice should decrease the data transmit power and the jammer should increase the jamming transmit power so that the covert communication requirement is satisfied.
It is worth noting that if we aim to guarantee  very high detection error probability at Eve  i.e., 99.9\%, the covert rate is quite limited.

\begin{figure}[t]
	\centering
	\includegraphics[width=4.2in,height=3.3in]{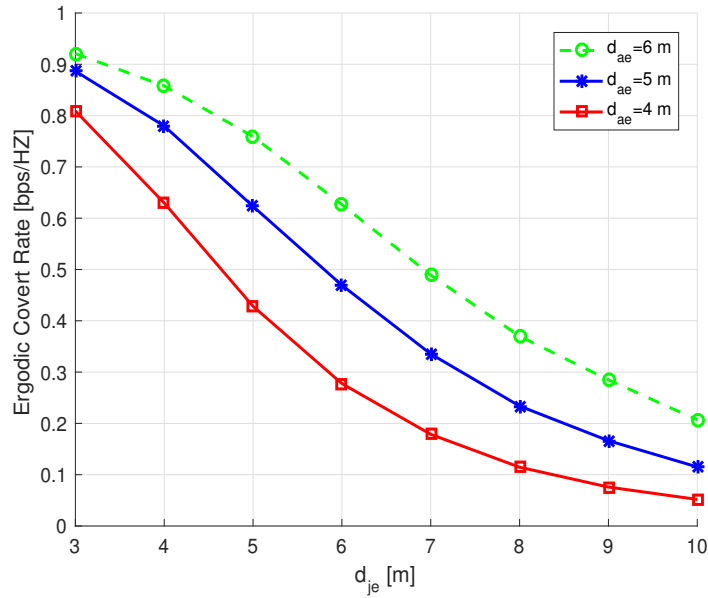} 
	\caption{\textcolor{black}{ Ergodic covert rate \emph{vs} $d_{je}$ and evaluation of the impact of $d_{ae}$ on covert rate, $\varepsilon=0.1$, $P_t=1.5$ Watts, $d_{ab}=5$ m.}}
	\label{d_jec}
\end{figure}
In Fig.~\ref{d_jec}, the covert rate versus the distance between the jammer and Eve is depicted. The distance between Alice and Eve is a parameter that is varied across curves.
As seen, the impact of the distance between Alice and Eve on  the covert rate is more than that of the distance between the jammer and Eve. 
\begin{figure}[t]
	\centering
	\includegraphics[width=4.2in,height=3.3in]{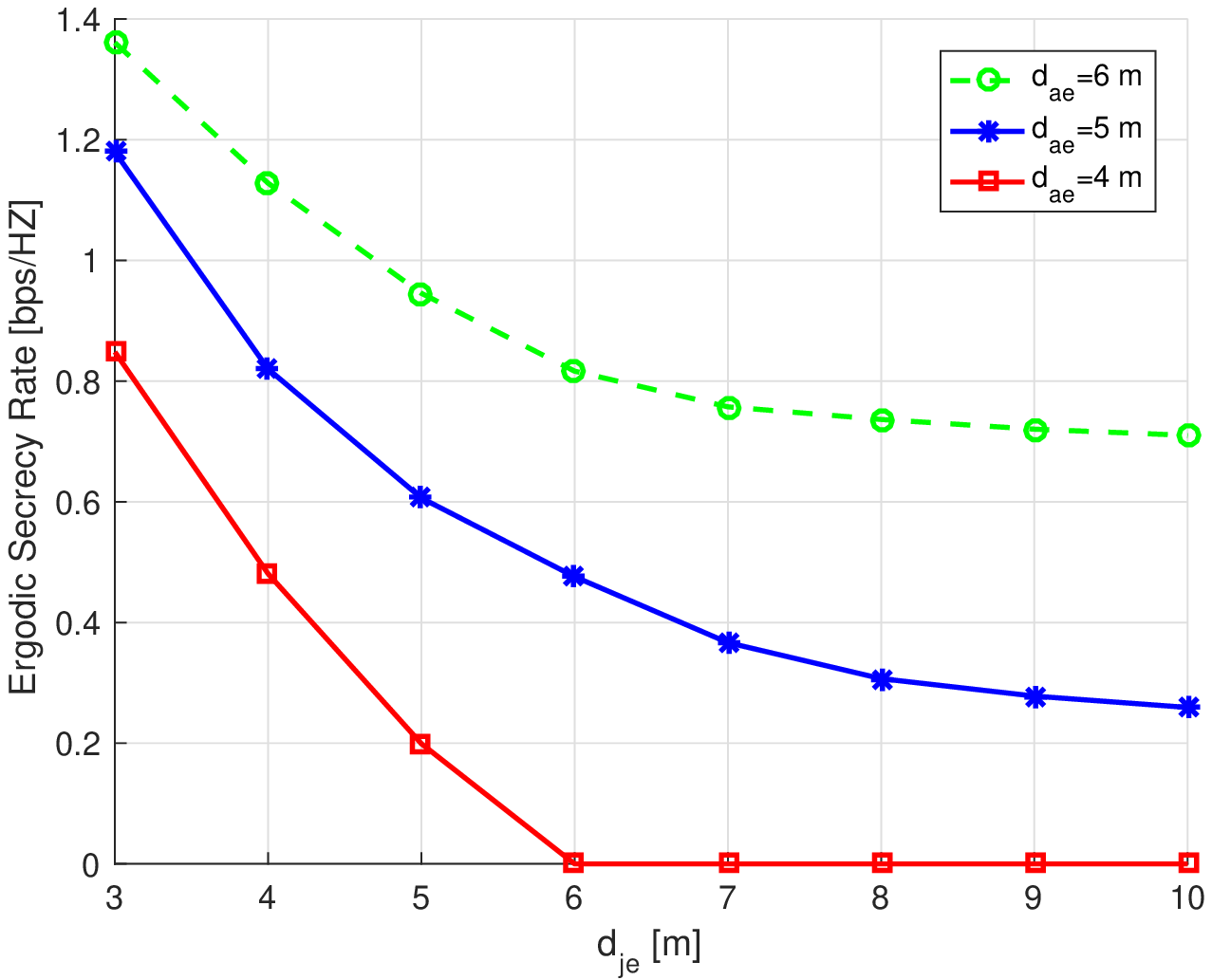} 
	\caption{\textcolor{black}{ Ergodic secrecy rate \emph{vs} $d_{je}$ and evaluation of the impact of $d_{ae}$ on covert rate, $P_t=1.5$ Watts, $d_{ab}=5$ m.}}
	\label{d_jes}
\end{figure}
In Fig.~\ref{d_jes}, the ergodic secrecy rate versus the distance between the jammer and Eve is shown. Furthermore, in this figure, we study the impact of the distance between Alice and Eve on the secrecy rate. 
This figure shows that the impact of distance of between Alice and Eve on secrecy rate is more than that of the distance between the jammer and Eve.

Finally, a comparison between Fig.~\ref{d_jec} and Fig.~\ref{d_jes} shows the sensitivity of information-theoretic security with respect to the location of Eve is more than that of covert communication.

\section{Conclusion}\label{Conclusion}
This paper investigated information-theoretic security and covert communication for a 4-node wiretap channel.
In order to make a comprehensive study, we investigated the power allocation problems for both methods. In this paper, we considered a practical assumption i.e., we assumed only CDI of channels are available for each of the secrecy and the covert communications scenario. 
Under a similar but not identical mathematical formulation, we formulated the optimization problems to maximize secrecy/covert rate subject to a transmit power constraint for both methods and a covert communication requirement for the covert communication method. Due to the non-convexity of the information-theoretic security optimization problem, we adopted the SCA method to solve the problem. For the covert communication optimization problem, we introduced an auxiliary variable to convert this non-convex optimization problem to a convex one.
Numerical results revealed that when Bob is near to Alice, the information theoretic security scheme has more rate than that of the covert
communication scheme but, perhaps surprisingly, this result is reversed when Bob is far from Alice. Moreover, we showed the sensitivity of information-theoretic security with respect to the location of Eve is more than that of covert communication.
In order to study the optimality gap, the performance of the proposed solution was compared with that obtained from the exhaustive method. The simulation results revealed that the optimality gap between the proposed solution method and the exhaustive search method is small.

\textcolor{black}{
	\section*{\sc Appendix B}
	Since Eve aims to minimize $\mathbb{P}_{FA} + \mathbb{P}_{MD}$, she does not select $\vartheta$ such that $\vartheta < \sigma_w^2$, as this would lead to $\mathbb{P}_{FA} + \mathbb{P}_{MD}=1$. Hence, we can consider the summation of probability of FA and MD $\vartheta > \sigma_w^2$ i.e., $\mathbb{P}_{FA} + \mathbb{P}_{MD}=1 + {e^{\frac{{ - \left( {\vartheta  - \sigma _e^2} \right)}}{{\left( {1 - \rho } \right){P_t}d_{je}^{ - \alpha }}}}} + \frac{A}{{\rho d_{ae}^{ - \alpha } - \left( {1 - \rho } \right)d_{je}^{ - \alpha }}}.$ Accordingly, to obtain $\vartheta ^*$, we set $\frac{{\partial \left(\mathbb{P}_{FA} + \mathbb{P}_{MD}\right)}}{{\partial \vartheta}}=0$, and by some mathematical manipulations, we find $
	{\vartheta ^*} = \left( {\frac{{\rho {P_t}d_{ae}^{ - \alpha }\left( {1 - \rho } \right){P_t}d_{je}^{ - \alpha }}}{{\left( {1 - \rho } \right){P_t}d_{je}^{ - \alpha } - \rho {P_t}d_{ae}^{ - \alpha }}}} \right)\ln \left( {\frac{{\left( {1 - \rho } \right){P_t}d_{je}^{ - \alpha }}}{{\rho {P_t}d_{ae}^{ - \alpha }}}} \right) + \sigma _e^2.
	$
}

\end{document}